\documentclass[article, twocolumn, longbibliography]{revtex4-1}
\usepackage{graphicx}              
\usepackage{hyperref}
\usepackage{amsmath}
\usepackage{natbib}
\usepackage{bm}
\usepackage{float}
\usepackage{color}
\usepackage{braket}
\usepackage{ulem}

\begin{document}
\title{Slow molecular beams from a cryogenic buffer gas source} 
\author{A. D. White}
\author{S. Popa}
\author{J. Mellado-Mu\~noz}
\author{N. J. Fitch}
\altaffiliation[Present address: ]{Infleqtion, 3030 Sterling Cir. Boulder CO 80301, USA}
\author{B. E. Sauer}
\author{J. Lim}
\author{M. R. Tarbutt}
\email{m.tarbutt@imperial.ac.uk}
\affiliation{Center for Cold Matter, Blackett Laboratory, Imperial College London, Prince Consort Road, London SW7 2AZ, United Kingdom}
\date{\today}   

\begin{abstract}
We study the properties of a cryogenic buffer gas source that uses a low temperature two-stage buffer gas cell to produce very slow beams of ytterbium monofluoride molecules. The molecules are produced by laser ablation inside the cell and extracted into a beam by a flow of cold helium. We measure the flux and velocity distribution of the beam as a function of ablation energy, helium flow rate, cell temperature, and the size of the gap between the first and second stages of the cell. We also compare the velocity distributions from one-stage and two-stage cells. The one-stage cell emits a beam with a speed of about 82~m~s$^{-1}$ and a translational temperature of 0.63~K. The slowest beams are obtained using the two-stage cell at the lowest achievable cell temperature of 1.8~K. This beam has a peak velocity of 56~m~s$^{-1}$ and a flux of $9 \times 10^9$ ground state molecules per steradian per pulse, with a substantial fraction at speeds below 40~m~s$^{-1}$. These slow molecules can be decelerated further by radiation pressure slowing and then captured in a magneto-optical trap.

\end{abstract}

\maketitle 

\section{Introduction}

The cryogenic buffer gas source~\cite{Maxwell2005, Hutzler2012} has become an important method for producing beams of atoms and molecules with high flux and low velocity. In this source, atoms and molecules of interest are formed inside a cell where they thermalize with cryogenically-cooled buffer gas, usually helium or neon, before exiting the cell to form a beam. This type of source is the starting point for all current experiments on magneto-optical trapping of molecules~\cite{Barry2014, Truppe2017b, Anderegg2017, Collopy2018, Lu2022, Vilas2022}, and has been used to load magnetic traps~\cite{Lu2014}, electric traps~\cite{Prehn2016}, Stark decelerators~\cite{Bulleid2012, Aggarwal2021}, centrifuge decelerators~\cite{Wu2017} and some atomic magneto-optical traps (MOTs)~\cite{Hemmerling2014, Lasner2021}. Cryogenic buffer gas sources are also used in experiments using molecules to measure the electric dipole moments of electrons and protons~\cite{Andreev2018, Fitch2020b, Grasdijk2021, Aggarwal2018}, studies of cold collisions~\cite{Wu2017, Koller2022}, spectroscopic studies~\cite{Santamaria2016} and the production of a wide variety of molecular beams including carbon clusters and organic molecules~\cite{Patterson2015, Straatsma2017}.

Cryogenic buffer gas sources usually operate in a flow regime intermediate between effusive and hydrodynamic. Often, the mean speed in the beam is around 150~m~s$^{-1}$. In some applications, lower velocities are highly prized. Important examples are experiments that aim to laser cool and trap heavy molecules or molecules that do not have completely closed optical cycling transitions. Laser cooling of these molecules is needed for tests of fundamental physics~\cite{Lim2018, Alauze2021, Augenbraun2020} and experiments with ultracold polyatomic molecules~\cite{Augenbraun2020b}, but is difficult because heavy molecules have large momentum and the momentum transfer is limited when there are leaks out of the cooling cycle. To produce lower velocity beams, a two-stage cell can be used~\cite{Patterson2007, Lu2011}. In this design, a molecular beam is extracted efficiently from a first-stage cell operating in the hydrodynamic regime and then enters a second ``slowing cell'' where the buffer gas density is much lower and collisions slow the beam without greatly reducing the flux. This method can reduce the beam velocity to about 40~m~s$^{-1}$. 

The properties of one-stage cryogenic sources have been studied in many papers, e.g.~\cite{Maxwell2005, Hutzler2011, Barry2011, Hutzler2012, Bulleid2013, Wright2023, Mooij2024arxiv}, but there is only one previous work studying two-stage sources in any detail~\cite{Lu2011}. Here, we characterize the properties of a two-stage cryogenic source of YbF molecules operating at 1.8~K, measuring how the velocity distribution depends on the main controllable parameters of the source. The beam has a flux of $9 \times 10^9$ ground state molecules per steradian per pulse with a peak velocity of 56~m~s$^{-1}$. About 12\% of these molecules have speeds below 40~m~s$^{-1}$, and about 3\% are below 30~m~s$^{-1}$. The beam is suitable for laser slowing and magneto-optical trapping. A trapped, ultracold sample of YbF would be an excellent resource for measuring the electron's electric dipole moment~\cite{Fitch2020b}, and laser cooling of YbF has been demonstrated~\cite{Lim2018, Alauze2021}. However, small leaks out of the laser cooling cycle~\cite{Zhang2022} hinder radiation pressure slowing, so starting from a very slow beam is a great advantage.  For example, YbF molecules moving at 37~m~s$^{-1}$ can be brought to rest by scattering $10^4$ photons, which is feasible. Alternatively, a slow beam could be used along with a Zeeman-Sisyphus decelerator so that only a few tens of photons need to be scattered~\cite{Fitch2016, Augenbraun2021}.

\section{Experiment Setup}
\label{sec:ex-setup}

\begin{figure*}[tb]
\center
\includegraphics[width=0.9\textwidth]{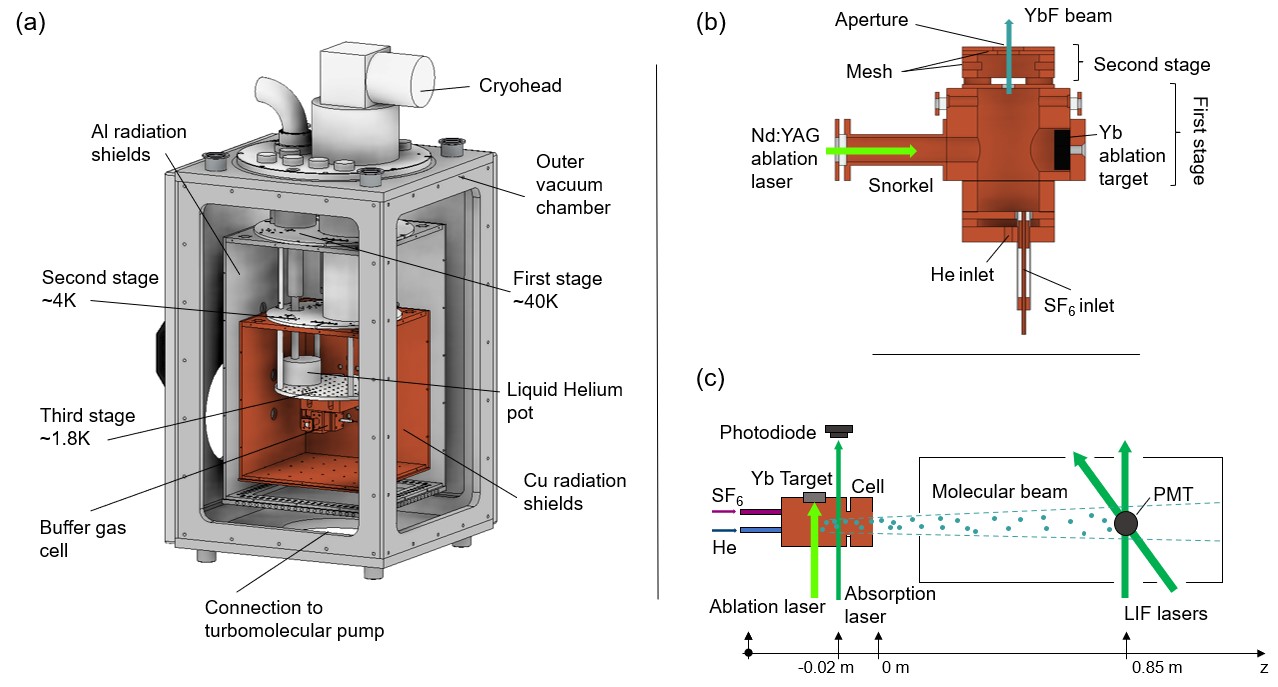}
\caption[Schematic of the cryogenic buffer gas source]{(a): Diagram of the molecular beam source. The vacuum chamber doors have been omitted. (b): Section through the two stage buffer gas cell. (c) Setup for characterising the source by absorption spectroscopy inside the cell and velocity-sensitive laser-induced-fluorescence spectroscopy in the beam.}
\label{fig:Source}
\end{figure*}

Figure \ref{fig:Source}(a) illustrates the cryogenic source. YbF molecules are produced inside a buffer gas cell by laser ablation of an Yb target in the presence of SF$_6$ gas. The molecules cool to low temperature through collisions with $^{4}$He gas that flows through the cell. The buffer gas cell is attached to the low-temperature stage of a three-stage cryocooler (ICEOxford). The base temperature is 1.1~K, but during operation the cell is at 1.8~K primarily due to the heat load from the ablation laser. A heater and thermistor mounted to the low-temperature stage are used to control the temperature of the cell. The low-temperature stage is inside a copper box attached to the second stage of the cryocooler which is cooled to 4~K. Charcoal glued to the inner surface of this box acts as a sorption pump, absorbing the helium that flows through the source. The temperature of the charcoal sorbs remain constant as the cell temperature is varied between 1.8 and 4.2~K, decoupling the influence of cell and sorb temperature on the properties of the beam. The copper box sits inside an aluminium box that is cooled to about 50~K by the first stage of the cryocooler and shields the lower temperature stages from room temperature radiation. Small holes in the two boxes allow the molecular beam to escape and provide optical access for the ablation laser and a probe laser used for absorption spectroscopy. The entire assembly is inside a vacuum chamber where the pressure is below 10$^{-8}$~mbar when the source is cold.

We use a two-stage cell following the design from Harvard~\cite{Lu2011} and illustrated in Fig.~\ref{fig:Source}(b). The cell is machined from oxygen-free copper and has a cylindrical bore of diameter 25.4~mm. The first stage has a length of 38~mm and an exit aperture of diameter $d_1 = 5$~mm. The second stage has a length of 9.5~mm and a 9~mm diameter exit aperture covered by a copper mesh with an open area of 37\%.  A variable gap between the two cells, typically set to 2.8~mm, controls the relative density of helium gas in the two stages. A pipe of diameter 3~mm delivers the helium to the cell. The helium is pre-cooled to 4~K by thermally anchoring the pipe to both the first and second stages. The gas enters the main body of the cell through a diffuser plate that helps to thermalize the gas with the cell walls. A second pipe of diameter 1.5~mm delivers SF$_6$ gas to the cell. This pipe is not in direct contact with any of the cryogenic surfaces, is thermally isolated from the cell by a teflon tube, and is maintained at 220~K using a heater attached to the pipe outside the radiation shield and a thermistor attached to the pipe close to the cell. The flow rates of He and SF$_6$ are controlled using mass flow controllers. The ablation target is a 9~mm diameter Yb disk attached to a removable plate inside the first stage of the cell. The target is polished with 400 grit sandpaper so that the surface is flat and uniform. This target is ablated by light from a Nd:YAG laser that produces pulses of wavelength 532~nm, duration 7~ns, maximum energy 125~mJ, and a repetition rate of 2~Hz. The light is focused onto the target with a spot size of 40~$\mu$m, and enters through a window offset from the cell to reduce coating by ablation products.  In the first-stage cell, the mean free path is much smaller than the size of the cell so the molecules produced by laser ablation become entrained in the flow of cold helium and thermalize with it. This is efficient in bringing the molecules into a beam and into the second-stage cell where the helium density and flow velocity are lower. Moreover, reflection from the mesh results in helium atoms travelling counter to the flow which helps to reduce the velocity of the molecules as they exit the cell. The result is a very slow molecular beam that still has a relatively high flux.

\begin{figure}[t]
\centering
\includegraphics[width=0.9\columnwidth, keepaspectratio]{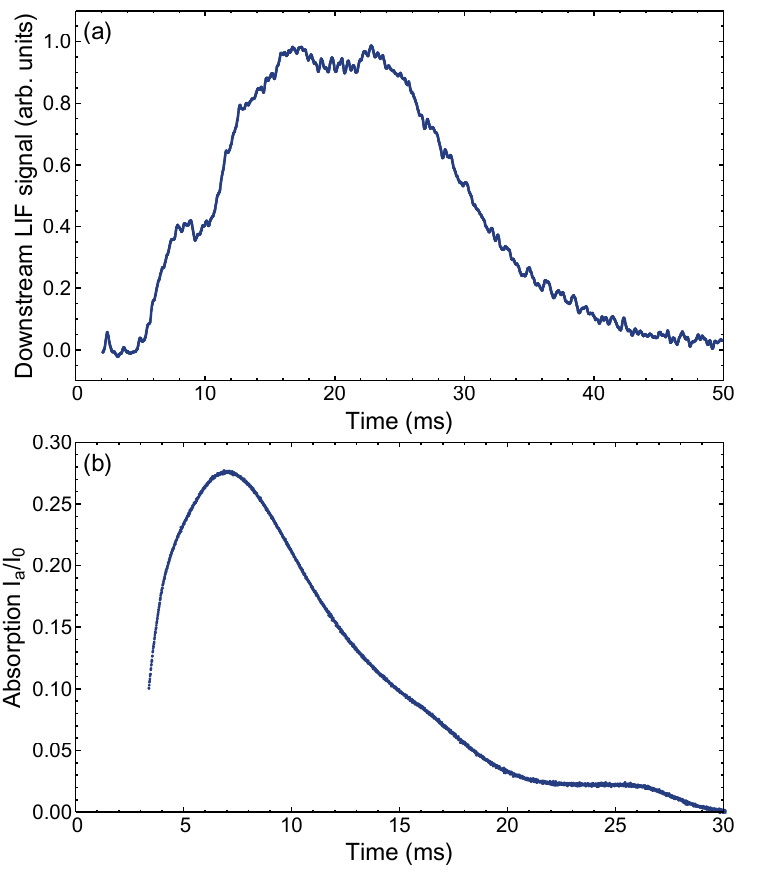}
\caption{\label{fig:tofs} Examples of fluorescence and absorption signals versus time for the two-stage cell with a gap size of 2.8~mm. Ablation energy is 55~mJ, helium flow is 2~sccm and cell temperature is 1.8~K.  (a) Laser induced fluorescence signal recorded on the PMT. (b) Laser absorption inside the cell. The signal at short times ($t < 3$~ms) is not shown because it is affected by electronic noise caused by the firing of the ablation laser.}
\end{figure}

\begin{figure*}[tb]
\centering
\includegraphics[width=0.75\textwidth, keepaspectratio]{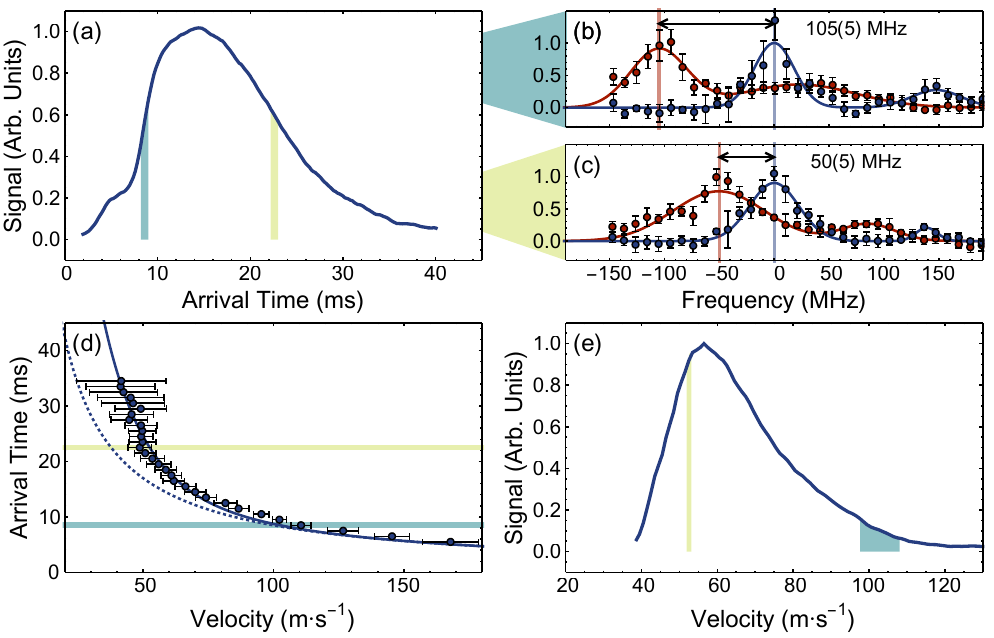}
\caption{\label{fig:VelocityMeasurement} Doppler sensitive laser-induced fluorescence spectroscopy to determine the longitudinal velocity distribution of molecules in the cryogenic buffer gas beam. (a) A time-of-flight profile is divided into equal time bins. (b,c) An angled/perpendicular probe laser is scanned across the P(1) resonance, shown in red/blue for two example time gates. The lines are fits to a sum of two Gaussians. (d) Points: arrival time ($t$) of molecules as a function of their velocity ($v$). Dashed line: fit to $t=d/v$. Solid line: fit to $t=d/v + t_0(v)$ where $t_0(v) = A e^{-\beta v}$. (e) Velocity distribution.}
\end{figure*}

We detect the molecules in the cell by absorption spectroscopy and detect the molecules in the beam by laser-induced fluorescence (LIF) spectroscopy, as illustrated in Fig.~\ref{fig:Source}(c). In both cases, a probe laser drives the P(1) component of the $\text{A}^{2}\Pi_{1/2}(v'=0) \leftarrow \text{X}^{2}\Sigma^{+}(v=0)$ transition of $^{174}$YbF, which has a wavelength of 552~nm and a linewidth of about 6~MHz. All hyperfine components of this transition are driven simultaneously by applying radio-frequency sidebands to the probe laser. The absorption and fluorescence are proportional to the number of molecules in the first rotationally-excited level ($N=1$) of ground state $^{174}$YbF, $\text{X}^{2}\Sigma^{+}(v=0)$. The fluorescence detector is situated 85~cm from the exit aperture of the cell. Here, the molecules can be probed by two laser beams, one at 90$^{\circ}$ and the other at 60$^{\circ}$ to the direction of the molecular beam. The velocity is determined from the Doppler shift between the two spectra produced by scanning the frequency of these probes. The probe beams have a diameter of 5~mm and a power of 7.5~mW, and the resulting fluorescence is imaged onto a photomultiplier tube (PMT). The collection optics consists of two lenses each of diameter 50~mm and focal length $f=43$~mm. The first lens is $f$ away from the centre of the beam, and the second is $f$ away from the PMT. The collection efficiency is doubled by reflecting the light emitted in the opposite direction using a mirror with a radius of curvature $R=50$~mm placed a distance $R$ from the beam. The probe power is high enough, and the interaction time long enough, that most molecules scatter photons until they decay into a higher-lying vibrational state. This means that the fluorescence signal has little dependence on the velocity. The flux of molecules is determined from the number of photons detected in each shot along with estimates of the solid angle of the detection volume, the solid angle for collecting photons, the PMT quantum efficiency and the number of photons that each molecule scatters.

\section{Results}

\subsection{Absorption and fluorescence signals}

Figure \ref{fig:tofs} shows examples of the fluorescence signal in the beam and the absorption signal in the cell, both as functions of the time since firing the ablation laser. The signals were recorded simultaneously. The peak of the absorption profile occurs at about 7~ms and then decays with a time constant which is also about 7~ms. This is consistent with the expected time for diffusion to the walls at the helium density used here~\cite{Skoff2011, PopaThesis}. The fluorescence signal is centred at 20~ms. Taking the distance between the two detection points (0.87~m) and dividing by the time between the absorption and fluorescence peaks (13~ms) gives a first estimate of the mean speed as 67~m~s$^{-1}$. At the peak of the absorption profile, about 28\% of the incident light is absorbed. Using a simple model of the absorption, and assuming the cell volume is filled uniformly, this corresponds to $1.7 \times 10^{11}$ $^{174}$YbF molecules in the cell (in the $N=1$ rotational level of the ground state)~\cite{PopaThesis}. This is a rough estimate. A more accurate measurement needs to account for saturation of the absorption and the effects of collisions and diffusion into the probe volume~\cite{Skoff2011}.

\subsection{Doppler velocimetry}

We have measured the flux and velocity distribution of the YbF beam as the key parameters of the source are varied. In the limit where the mean free path inside the cell is very short such that that the YbF is fully entrained in the helim flow, we expect the YbF velocity to reach that of a supersonic beam of helium from a reservoir at 1.8~K, which is 137~m/s. We also expect these conditions to give high flux because the molecules are swept out of the cell efficiently. In the opposite limit where the mean free path is long compared to the diameter of the exit aperture, we reach an effusive source of YbF where the mean speed is only 17~m/s and the flux is much lower. Since the source operates between these two regimes, and since collisions with helium will always tend to boost the YbF speed, we expect a speed between these two limits.

Although a velocity distribution could be determined directly from the time-of-flight profile, this is not accurate when the time taken to exit the cell is a significant fraction of the total time, as it is here. A velocity distribution could also be determined by integrating over all arrival times and comparing the spectra recorded using the two probe lasers. In the ideal situation where the spectrum recorded using the 90$^{\circ}$ probe is a single, narrow line, the spectrum recorded with the angled probe directly measures the velocity distribution. However, our spectrum has hyperfine structure on a similar scale to Doppler broadening, which can skew the inferred velocity distributions. Furthermore, the signal-to-noise ratio of the spectral data is not as good as the time-of-flight profile where all the data is taken on resonance. Instead, we reconstruct the velocity distribution by using the time-of-flight profile and the Doppler shift data together. 

The method is illustrated in Fig.~\ref{fig:VelocityMeasurement} and is similar to the one described in Ref.~\cite{Truppe2017}.
For each time bin in the time-of-flight (TOF) profile (Fig.~\ref{fig:VelocityMeasurement}(a)), we measure the Doppler shift between the spectra recorded using the two probe lasers. Two examples of these spectra from different time bins are shown in Fig.~\ref{fig:VelocityMeasurement}(b,c)). The probe laser has rf sidebands applied in order to address the four hyperfine components of the transition. When the laser is tuned so that all four components are driven, we see a large peak in the spectrum. There are also smaller peaks on both the high- and low-frequency sides, where the frequency components of the light drive a subset of the hyperfine components of the molecule. When probed at 90$^{\circ}$, these smaller peaks are well resolved. The spread of velocities in a single time bin is narrow enough that this hyperfine structure remains resolved in the Doppler-sensitive spectrum. In Fig.~\ref{fig:VelocityMeasurement}(b,c), only the large peak and the smaller peak on the low-frequency side are visible, so we fit the data to a sum of two Gaussian profiles.

From the Doppler shifts between the two spectra we obtain the mean longitudinal velocity, $v$, as a function of arrival time, $t$, which is shown by the points in Fig.~\ref{fig:VelocityMeasurement}(d). We model the relation between $t$ and $v$ as $t = d/v + t_0(v)$, where the first term is the time for free flight over the distance $d=85$~cm from the exit of the cell to the detector, and the second term is the time taken for molecules of speed $v$ to exit the cell. Since we measure $t(v)$, the data yield $t_0(v)$. To produce a smooth function, we fit this to the model $t_0(v) = A e^{-\beta v}$, where $A$ and $\beta$ are free parameters. We do not have a physical basis for this particular model, but find that it tends to fit well, as can be seen by the solid line in Fig.~\ref{fig:VelocityMeasurement}(d). This is sufficient for our purpose, since our only concern is to represent the discrete data by a smooth function, and since $t_0(v)$ is only a small part of $t(v)$. Note that a model that neglects the exit times does not fit well to the data, as shown by the dashed line in the figure. To construct the velocity distribution, we take velocity bin $v$ of width $\delta v$, determine the width $\delta t$ of the equivalent time bin using the fit, count the number of molecules arriving within this time interval, and assign them all to this velocity bin. This is an approximate procedure that works well when the range of velocities in each time bin is small. Figure \ref{fig:VelocityMeasurement}(e) shows the velocity distribution determined this way. It peaks at 53~m~s$^{-1}$ and has a full width at half maximum (FWHM) of 28~m~s$^{-1}$.

\subsection{One-stage and two-stage cells} 

\begin{figure}[t]
	\begin{center}
		\includegraphics[width=0.9\columnwidth]{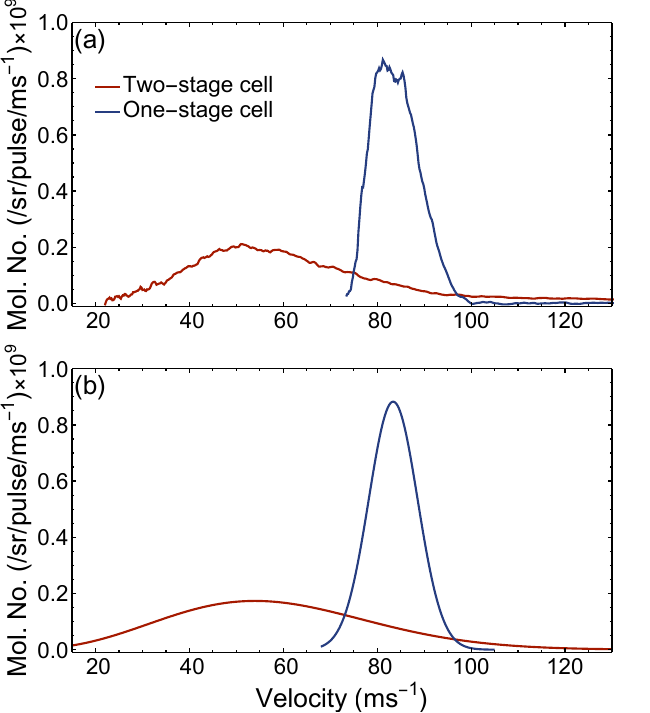}
	\end{center}
	\caption{(a) Measured velocity distributions for the one-stage and two-stage cells. Ablation energy is 49~mJ and helium flow rate is 2~sccm. For the two-stage cell the gap size is 2.8~mm. (b) Result of fitting Eq.~(\ref{eq:fss}) to the one-stage data and Eq.~(\ref{eq:feff}) to the two-stage data.}
	\label{fig:2stagevs1stage}
\end{figure}

Figure \ref{fig:2stagevs1stage}(a) shows velocity profiles measured for the one-stage and two-stage cells when the ablation energy is 49~mJ and the helium flow rate is 2~sccm. The distribution from the one-stage cell peaks at 81~m~s$^{-1}$ and has a FWHM of only 12~m~s$^{-1}$. Such a narrow distribution is characteristic of a supersonic beam. It is common to model the velocity distribution of a supersonic beam as
\begin{equation}
    f_{\rm ss}(v) = A v^3\exp\left(-\frac{M(v-v_0)^2}{2 k_{\rm B} T_{\rm trans}}\right),
    \label{eq:fss}
\end{equation}
where $A$ is a normalization constant, $M$ is the mass of the molecule, $v_0$ is the central velocity and $T_{\rm trans}$ is the translational temperature of the beam. The blue distribution shown in Fig.~\ref{fig:2stagevs1stage}(b) is the best fit of the one-stage data to this model. The best fit parameters are $v_0 = 82.4 \pm 0.1$~m~s$^{-1}$ and $T_{\rm trans} = 0.63 \pm 0.01$~K. The translational temperature is below the cell temperature, which is typical for supersonic beams which cool as they expand from the higher-pressure region of the cell to the low pressure region beyond. The production of such a slow beam with such a narrow velocity distribution is unique and remarkable. We have checked the width of the distribution by examining the widths of the spectral features (like those in Fig.~\ref{fig:VelocityMeasurement}(b,c)) after integrating over the entire time-of-flight profile (instead of taking narrow time windows). The rms width is $16.9 \pm 0.6$~MHz for the 90$^\circ$ probe and $16.8 \pm 0.7$~MHz for the 60$^\circ$ probe, showing that there is no measurable Doppler broadening. The uncertainties on this measurement translate to an upper limit of 15.6~m~s$^{-1}$ for the FWHM of the velocity distribution, which is not much larger than the 12~m~s$^{-1}$ we measure.

A fully supersonic beam of a monoatomic carrier gas would have a speed of $v_0 = \sqrt{5 k_{\rm B}T/m}$, where $T$ is the cell temperature and $m$ is the mass of the carrier gas. For He at 1.8~K, we would expect $v_0 = 137$~m~s$^{-1}$, considerably higher than observed.  For the flow of 2~sccm used here, the helium density in the one-stage cell, $n_{\rm He,1}$, is in the range $(1.0-1.8)\times 10^{21}$~m$^{-3}$, depending on the flow regime~\cite{Bulleid2013}. Using the cross section for He-He collisions measured at 2.1~K, $\sigma_{\rm He} = 1.44\times 10^{-18}$~m$^2$~\cite{Oates1971}, we find a mean free path $\lambda_{\rm 
 He,1}=1/(\sqrt{2}n_{\rm He,1}\sigma_{\rm He})$ in the range 0.27-0.49~mm. The flow regime is often characterized by the Reynolds number, $R \approx 2d_1/\lambda_{\rm He,1}$~\cite{Hutzler2012}. When $R<1$ the beam is effusive, when $R>100$ it is in the supersonic regime, and for intermediate values it is in the partially hydrodynamic regime. Taking a value of $\lambda_{\rm He,1}$ in the middle of our range, we find $R \approx 26$ which puts the flow in the partially hydrodynamic regime. This is consistent with our observation that there is translational cooling of the molecules as in a supersonic expansion, but that the molecules have not reached the velocity of a fully supersonic beam. 
 
The addition of the second stage reduces the peak of the velocity distribution to 56~m~s$^{-1}$ and increases its width to 38~m~s$^{-1}$, showing that the two-stage cell produces a more effusive beam. There is a substantial flux of molecules with speeds below 30~m~s$^{-1}$. We model this near-effusive beam, with the velocity distribution 
\begin{equation}
    f_{\rm eff}(v) = B v^3\exp\left(-\frac{m^* v^2}{2 k_{\rm B} T}\right),
    \label{eq:feff}
\end{equation}
where $B$ is a normalization constant, $T$ is the cell temperature and $m^*$ is an effective mass. The red distribution in Fig.~\ref{fig:2stagevs1stage}(b) is the best fit of the two-stage data to this model. The fit gives $m^*= 15.4 \pm 0.1$~u, reflecting the observation that the beam is slower than an effusive beam of He but faster than an effusive beam of YbF, and implying that there are still collisions between the two species in the vicinity of the exit aperture. We have also measured the rotational temperature of the beam produced by the two stage cell by measuring the relative intensities of several Q lines in the spectrum. We found a temperature of $1.0 \pm 0.1$~K, a little below the cell temperature of 1.8~K, implying some rotational cooling due to collisions close to the exit aperture.

Remarkably, the two sources produce similar fluxes in $N=1$: $(1.1 \pm 0.1)\times 10^{10}$ molecules/sr/pulse for the one-stage cell and $(9.3 \pm 0.9) \times 10^{9}$ molecules/sr/pulse for the two stage cell. Note however that the rotational temperature could be different for the one-stage and two-stage cells (we only measured it for the two-stage cell).

Having compared the one- and two-stage cells, the rest of this paper focusses on the properties of the beam produced by the two-stage cell. Unless stated otherwise, the cell temperature is 1.8~K and the size of the gap between the first and second stages is 2.8~mm.

\subsection{Exit times}

\begin{figure}[t]
\centering
\includegraphics[width=\columnwidth, keepaspectratio]{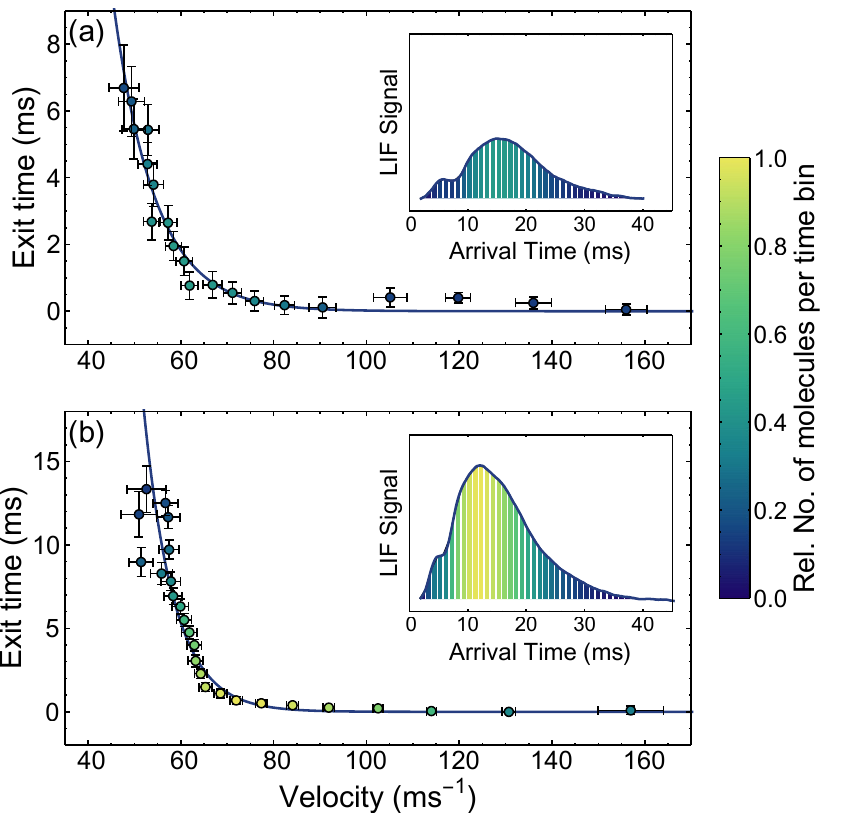}
\caption{\label{fig:PhaseSpace} Exit time from the two-stage cell versus velocity. Solid lines are fits to $t_0(v)=A e^{-\beta v}$. (a) Ablation energy: 69~mJ, helium flow: 8~sccm. (b) Ablation energy: 88~mJ, helium flow: 10~sccm. Insets show the corresponding time-of-flight profiles and colour indicates the relative number of molecules in each time bin.
}
\end{figure}

\begin{figure*}[t]
\centering
\includegraphics[width=0.98\textwidth, keepaspectratio]{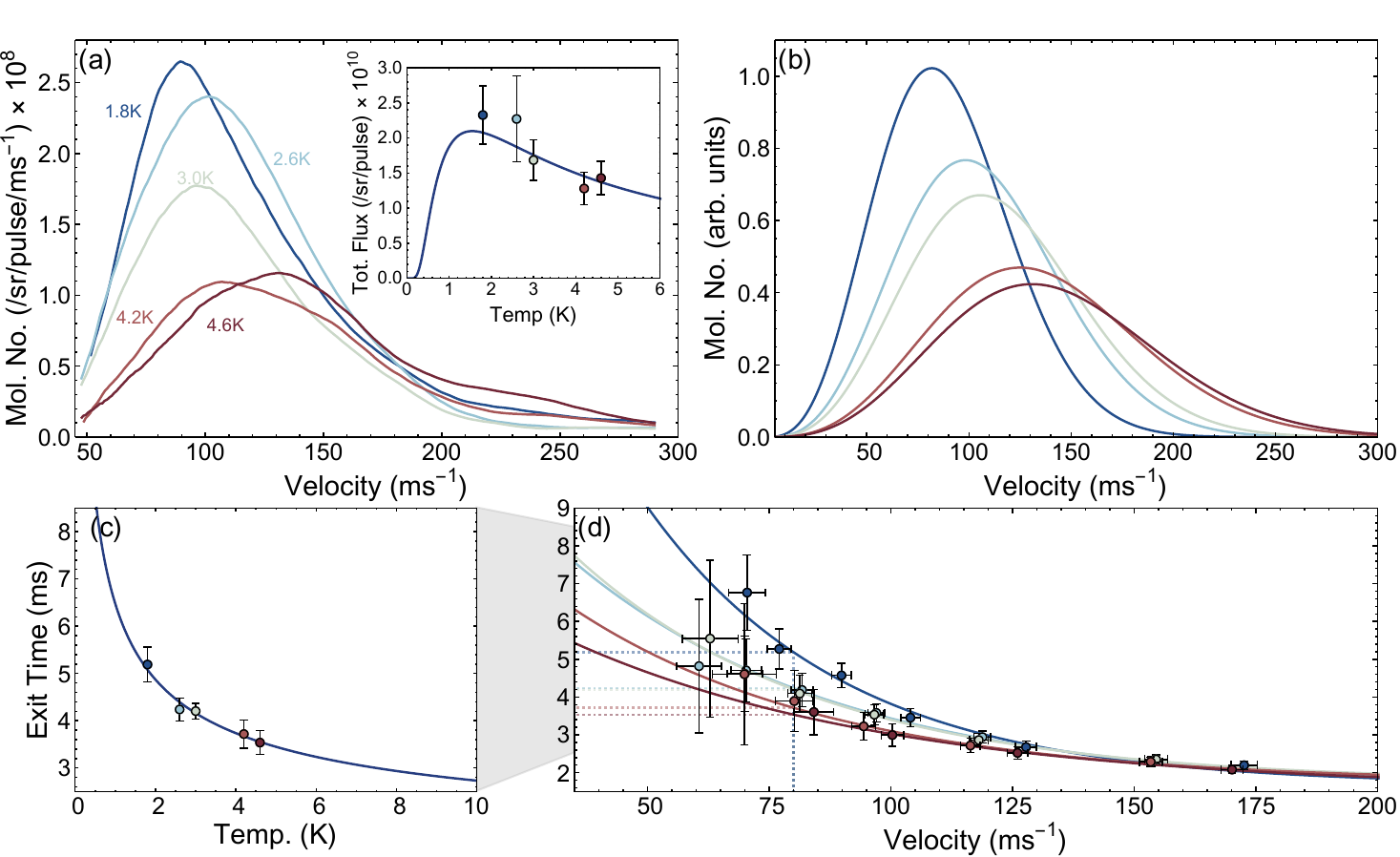}
\caption{\label{fig:TempDependence} Temperature dependence of velocity distributions. (a) Measured YbF velocity distributions for various cell temperatures, $T$. Inset, points: total molecular flux. Inset, line: fit to a Boltzmann distribution for the population in $N=1$. (b) Modelled velocity distributions at each cell temperature, assuming an effusive beam with an effective mass of $m^*=6.7$~u. (c) Points: Exit times as a function of temperature for molecules travelling at 80~m/s. Line: Fit to $t_0 = t_{00} + A/\sqrt{T} $. (d) Points: Exit times as a function of velocity for different cell temperatures. Lines: Single exponential decays.  
}
\end{figure*}

Figure \ref{fig:PhaseSpace} shows exit times as a function of velocity, $t_0(v)$, for two distinct sets of parameters, together with the time-of-flight profiles shown as insets. The colour coding indicates the number of molecules at each velocity or arrival time. In Fig.~\ref{fig:PhaseSpace}(a) the ablation energy is 69~mJ and the helium flow is 8~sccm, which produces a velocity distribution peaked at 63~m~s$^{-1}$. In Fig.~\ref{fig:PhaseSpace}(b) the ablation energy is increased to  88~mJ and the helium flow increased to 10~sccm, producing a more intense beam with a peak velocity at 80~m~s$^{-1}$. For the slow molecules, the exit time correlates strongly with speed with the slower molecules exiting later. Molecules with speeds below about 60~m~s$^{-1}$ have characteristic exit times of a few ms, consistent with the typical decay time of the in-cell density determined by absorption measurements (see Fig.~\ref{fig:tofs}(b)). The exit times are systematically longer at the higher flow rate. This is because higher flow gives higher helium density in the cell, resulting in more atom-molecule collisions and a longer time for molecules to diffuse to the exit of the cell. Molecules travelling faster than 65~m~s$^{-1}$ leave the cell quickly ($t_0 < 1$~ms) and there is little correlation between the velocity and the exit time. As noted above, we fit the exit time data to the model $t_0(v)=A e^{-\beta v}$, which seems to fit well. 

\subsection{Cell temperature}

Figure \ref{fig:TempDependence}(a) shows velocity distributions measured for the two-stage cell at five different cell temperatures. For these measurements, the helium flow is 2~sccm and the ablation energy is 80~mJ, resulting in faster beams than the one shown in Fig.~\ref{fig:2stagevs1stage}. Lowering the temperature shifts the peak of the velocity distribution to lower values, increases the total flux of molecules in $N=1$, and dramatically increases the flux at low velocity. This illustrates the great benefits of cooling the cell to lower temperatures for the purpose of producing very slow molecular beams. The inset in Figure \ref{fig:TempDependence}(a) shows the total flux of molecules in $N=1$ determined by integrating the velocity distributions. The line is a fit to the population expected in $N=1$ according to a Boltzmann distribution. This fits well, implying that the total flux across all rotational states is independent of cell temperature. Figure \ref{fig:TempDependence}(b) shows velocity distributions calculated for effusive beams at each of these cell temperatures, where we have fixed the effective mass to $m^*=6.7$~u in all cases and assumed the population in $N=1$ follows the Boltzmann distribution for each cell temperature. We see that this simple model provides a reasonably good description of the velocity distribution for all the cell temperatures studied. The choice of $m^*$ is the one that gives the best match to the data. The lower value of $m^*$ found here, relative to that found for Fig.~\ref{fig:2stagevs1stage}, reflects the higher velocity obtained at higher ablation energies, $m^*$ being the only parameter in Eq.~(\ref{eq:feff}) that affects the mean speed when $T$ is fixed.

Figure \ref{fig:TempDependence}(d) shows the exit times as a function of velocity at each temperature. As before, we fit single exponential decays to the data. As the cell temperature increases, the exit times get shorter. This can be seen more clearly in Fig.~\ref{fig:TempDependence}(c) which plots the exit time as a function of temperature for molecules travelling at 80~m~s$^{-1}$. The time taken to extract molecules from the cell should scale inversely with the velocity of the buffer gas, which in turn is proportional to the square root of the cell temperature, $T$. Thus, we fit the model $t_0 = t_{00} + A/\sqrt{T} $ to the data. This fits well, as shown by the line in Fig.~\ref{fig:TempDependence}(c).

\subsection{Gap size}

\begin{figure}[t]
	\begin{center}
             \includegraphics[width=0.88\columnwidth]{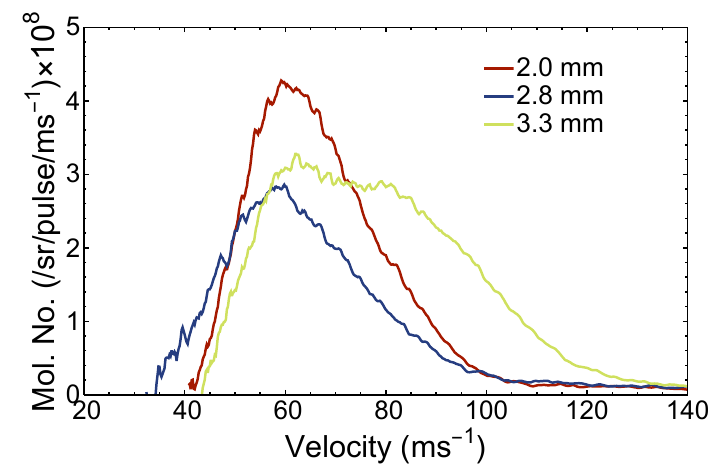}
	\end{center}
	\caption{Velocity distributions for three different gaps between the 
 first and second stages. Ablation energy is 49~mJ, helium flow is 2~sccm, and cell temperature is 1.8~K.}
	\label{fig:gaps}
\end{figure}

\begin{figure*}[t]
\centering
\includegraphics[width=0.9\textwidth, keepaspectratio]{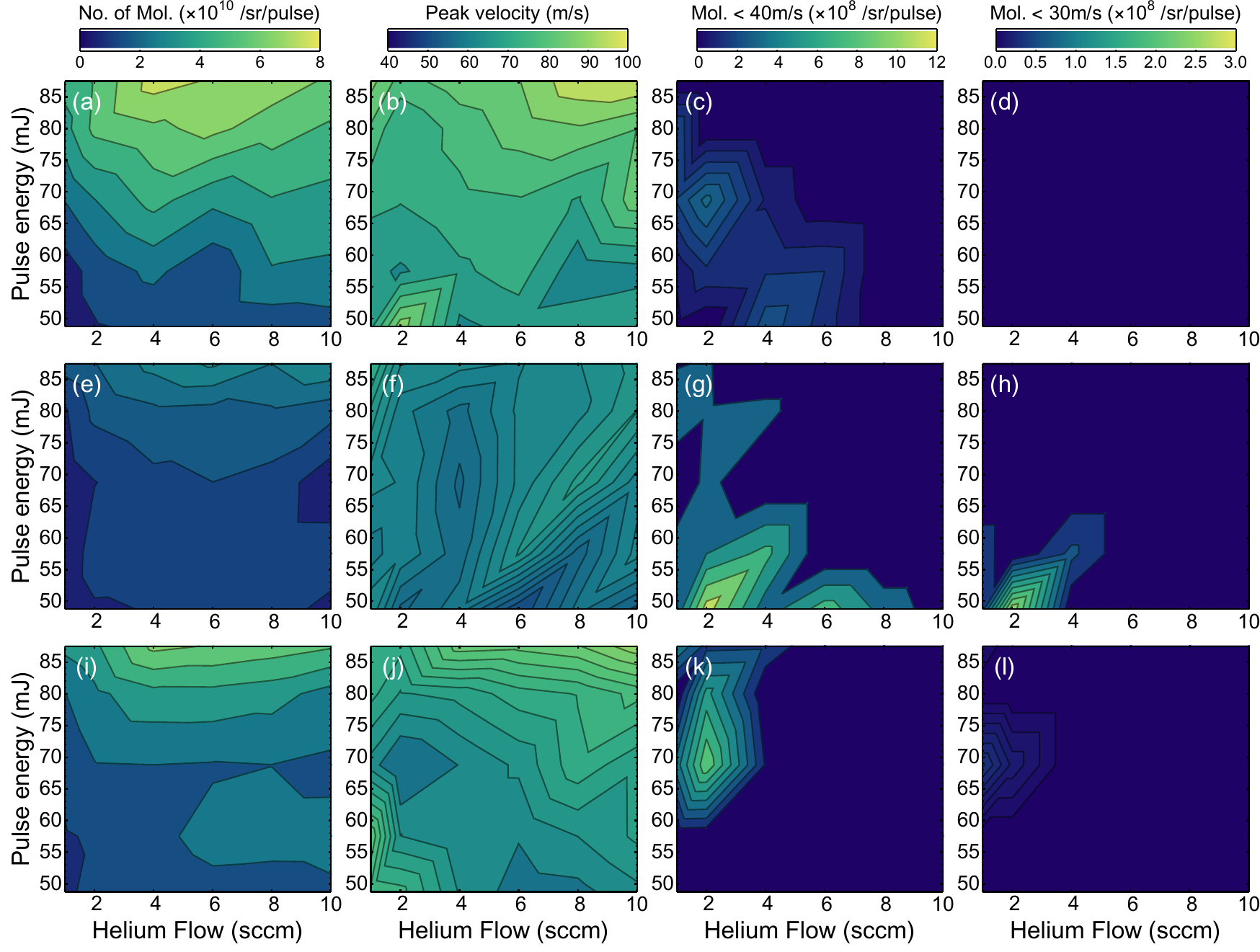}
\caption[Molecular beam characteristics as a function of ablation pulse energy and helium flow rate]{\label{fig:ParameterMap} Molecular flux, peak velocity and flux of molecules with speeds below 40~m/s and 30~m/s, as a function of ablation pulse energy and helium flow rate. (a)-(d) 2.0~mm gap between two stages of the cell.
(e)-(h) 2.8~mm gap. (i)-(l) 3.3~mm gap.
}
\end{figure*}

The parameters of the beam depend strongly on the size of the gap between the first and second stages of the cell. Helium flows out of this gap, so the gap size is a parameter that controls the relative helium density in the two cells. A smaller gap gives a higher helium density, and thus a shorter mean free path, in the second stage. Figure \ref{fig:gaps} shows the velocity profiles for three different gap sizes. The slowest beams are formed when the gap is 2.8~mm. With a smaller gap (2~mm) the helium beam tends more towards the supersonic regime and the molecules are partly entrained in this helium flow, giving a faster beam with a narrower velocity distribution. At larger gap sizes (3.3~mm) there are not enough collisions in the second stage to slow down the molecules coming out of the first stage, again giving a faster beam. Note that, of the three gaps used, the 2.8~mm gap gives the lowest total beam flux, but gives the highest flux of molecules below 40~m~s$^{-1}$. Thus, the best choice of gap depends on the quantity to be optimised. The data shown in Fig.~\ref{fig:gaps} is for a particular choice of ablation energy and helium flow, but we found that these general trends hold for all flows between 1 and 10~sccm and all ablation energies between 49 and 88~mJ.

\subsection{Optimization} 
\label{sec:optimization}

For the two-stage cell, we have measured velocity distributions across a wide range of parameters: helium flows in the range 1-10~sccm, ablation energies in the range 49-88~mJ and three gap sizes from 2.0 to 3.3~mm. The SF$_{6}$ flow rate has only a minor effect on the velocity distribution, so was fixed at 0.005~sccm which was found to maximise the YbF flux. Figure \ref{fig:ParameterMap} summarizes the results of these measurements. The figure shows the total flux in $N=1$, the peak velocity, and the flux of molecules below 40~m~s$^{-1}$ and below 30~m~s$^{-1}$, all as functions of ablation energy and helium flow, and for three different gap sizes.  For the 2.0~mm gap, the total flux and the peak velocity both increase with ablation energy. The flux does not depend strongly on the helium flow, whereas the peak velocity tends to increase with helium flow. The flux of molecules below 40~m~s$^{-1}$ is only substantial at relatively low flows and ablation energies, and there are no molecules below 30~m~s$^{-1}$ for this gap size. For the 2.8~mm gap the total flux has only a weak dependence on helium flow and is relatively constant for ablation energies up to 70~mJ, but then increases for higher energies. The peak velocity has little dependence on the ablation energy but tends to increase gradually with flow. At low ablation energy and flows close to 2 sccm there is a substantial flux with speeds below 40~m~s$^{-1}$, and even below 30~m~s$^{-1}$. The trends for the 3.3~mm gap are similar to those for the 2.8~mm gap. The total flux can be larger, but the peak velocity is significantly higher. There is a substantial flux below 40~m~s$^{-1}$, and a very small flux below 30~m~s$^{-1}$, when the ablation energy is 70~mJ and the flow is 2~sccm. To maximize the flux at the lowest velocities, we find it best to use a gap of 2.8~mm, a flow of 2~sccm and an ablation energy of 50~mJ. If a higher threshold velocity is acceptable, the optimum tends to move to higher flows and ablation energies. For example, to maximize flux below 50~m~s$^{-1}$, the best parameters are 4~sccm and 80~mJ.

\subsection{Stability}

The properties of the source tend to change over time due to ageing of the ablation target, ablation dust coating the cell walls, and contamination of the source. We typically ablate the same spot on the target for 2-3 hours. During this time the flux of molecules gradually decays, but it can be recovered by moving to a new spot on the target. Some spots are better than others. We tend to find that the total flux decreases and the peak velocity increases on a timescale of 2-3 weeks. Source parameters are typically recovered when the source is opened, cleaned and re-assembled. Some of the studies reported here (e.g. changing gap size or swapping between one-stage and two-stage cells) required a complete disassembly of the source, and source properties are not fully reproducible between one assembly and the next. Nevertheless, we believe all trends reported in this paper are reproducible.

\section{Conclusions}

We have characterized the performance of both a one-stage and a two-stage cryogenic buffer gas source across a wide range of parameters. The one-stage source emits a beam travelling at about 82~m~s$^{-1}$ with a translational temperature of only 0.63~K. Such a slow and cold velocity distribution is very well suited for injection into a decelerator, such as a travelling-wave Stark or Zeeman decelerator~\cite{Lavert-Ofir2011, Bulleid2012, Aggarwal2021}. The two-stage source is versatile and can be optimized either for high flux or for low speed. In the first case, the peak speed is about 95~m~s$^{-1}$ and the flux is about $7.5 \times 10^{10}$ molecules per steradian per pulse in $\text{X}^{2}\Sigma^+ (v=0, N=1)$. This flux is the same as reported from a one-stage cell at about 3K~\cite{Wright2023}, but the speed is roughly a factor two lower. In the second case the velocity is reduced to 56~m~s$^{-1}$ and the flux is $9 \times 10^9$ per steradian per pulse. About 12\% of these molecules have speeds below 40~m~s$^{-1}$ and about 3\% below 30~m~s$^{-1}$. The presence of a substantial flux of molecules with very low speeds is a major advantage for some applications, especially for radiation-pressure slowing of the beam down to the capture velocity of a magneto-optical trap (MOT), which is estimated to be 10~m~s$^{-1}$ \cite{Fitch2021b}. Molecules that start at 40~m~s$^{-1}$ will reach this capture velocity after scattering $8\times 10^3$ photons from a counter-propagating laser.

The optimum choice of ablation energy, helium flow, and gap size depends on the desired speed, and can be read from the maps presented in Fig.~\ref{fig:ParameterMap}. As the temperature of the cell is lowered, the beam velocity decreases significantly, but the total flux in all rotational states is approximately constant. The population in the lowest rotational state increases as the cell temperature is lowered, due to cooling of the rotational motion. For making the slowest possible beams, and driving more of the population to the ground state, a further reduction in the cell temperature would be beneficial. With no heat load from the ablation laser, the temperature falls to 1.1~K, so a method to produce the same ablation yield with lower ablation energies is desirable. Alternatively, a reduction in heat load per ablation pulse could be used to run the source at higher repetition rate, currently limited to 2~Hz. In this context, it would be interesting to study the dependence on ablation wavelength and pulse duration. A cryocooler offering even lower base temperatures, or greater cooling power, would be desirable. At the typical buffer gas densities used (of order $10^{21}$~m$^{-3}$), $^{4}$He remains a gas down to about 0.6~K. If a temperature below this could be obtained, $^{3}$He buffer gas might be needed.

The velocity distribution depends very strongly on the size of the gap between the first and second stages of the cell. In the present work, we only studied three gap sizes because warming up, re-configuring the cell, and cooling down again takes three days. A method for adjusting the gap size in situ would be very beneficial, allowing for much finer tuning of this important parameter. We did not study the dependence on other geometrical factors such as the cell volume, aperture size and mesh transparency. The formation of slow-moving high-flux beams requires enough collisions to thermalize the molecules and sweep them out efficiently, but not so many to boost the speed. This is a delicate balance, so it is likely that changing any one of these geometrical parameters will substantially change the beam properties, requiring a re-optimization of ablation energy, helium flow and gap size. Computational modelling~\cite{Takahashi2021} of a two-stage cryogenic source is likely to be very helpful in building a better understanding of these sources and how they depend on these geometric factors.

We have not measured the divergence of the molecular beam, but we suppose that it is similar to the divergence from other cryogenic sources where the typical solid angle is 0.3~sr~\cite{Barry2011, Hutzler2011, Skoff2011}. Combining this with our estimates of the number of molecules produced inside the cell and the downstream flux, we estimate that the fraction of molecules extracted into the beam is 1.6\% when optimized for low speed, rising to 13\% when optimized for high flux. Due to the substantial divergence, the vast majority of the molecules exiting the cell never reach the laser-induced fluorescence detector and could not be captured in a MOT either. This divergence can be reduced enormously using a region of transverse laser cooling close to the source to produce a much brighter beam~\cite{Alauze2021}. A further increase in brightness is predicted by using a magnetic lens to focus molecules from the source into a region of transverse laser cooling~\cite{Fitch2021b}. For a further increase in useable flux, population in multiple rotational and hyperfine states can be optically pumped into a single rotational and hyperfine state using the combination of an optical pumping laser and microwaves to couple rotational states together~\cite{Ho2020}. In this way, most of the available population can be utilized.

While we have focussed on YbF, our results are unlikely to depend on any details of the species other than the mass. The same methods can be applied to a wide range of other molecules and are likely to be most useful in molecule-based tests of fundamental physics~\cite{Safronova2018} and for laser cooling and trapping complex species with party-closed optical cycling transitions~\cite{Ivanov2020, Zhu2022}.

\section*{Acknowledgements}
We are grateful to John Doyle and his group for providing the design of the buffer gas cell and information about how to use it, and are grateful to Stefan Truppe for many helpful discussions about buffer gas sources. We thank the whole `LatticeEDM' team for helpful discussions and for making additional investigations to support this work. We thank Jon Dyne and David Pitman for their expert technical support. This research has been funded in part by the Gordon and Betty Moore Foundation through Grants 8864 \& GBMF12327 (DOI 10.37807/GBMF12327), and by the Alfred P. Sloan Foundation under Grants G-2019-12505 \& G-2023-21035, and by UKRI under grants EP/X030180/1, ST/V00428X/1 and ST/Y509978/1.

\bibliography{references}

\end{document}